**Coupling between COVID-19 and seasonal influenza leads to synchronization of their dynamics**

Jorge P. Rodríguez[1*], Víctor M. Eguíluz[2]

1. Non-affiliated researcher, Palma de Mallorca (Spain).
2. Instituto de Física Interdisciplinar y Sistemas Complejos IFISC (CSIC-UIB), Palma de Mallorca (Spain).
* Correspondence should be addressed to jorgeprodriguezg @ gmail . com

**Interactions between COVID-19 and other pathogens may change their dynamics. Specifically, this may hinder the modelling of empirical data when the symptoms of both infections are hard to distinguish. We introduce a model coupling the dynamics of COVID-19 and seasonal influenza, simulating cooperation, competition and asymmetric interactions. We find that the coupling synchronizes both infections, with a strong influence on the dynamics of influenza, reducing its time extent to a half.**

The COVID-19 outbreak, declared as a pandemic on March 11, 2020 by the World Health Organization, is challenging the scientific community, from the forecast and design of non-pharmaceutical interventions to the fast development of efficient treatments and vaccines. In fact, at the end of July 2020, there were more than 17 million confirmed cases around the world, which has led to more than 0.6 million deaths, according to the global data platform developed by John Hopkins University.

One of the uncertainties about COVID-19 is how this infection interacts with other infections in the hosts. Interactions can be neutral, such that the infections are independent and they do not modify mutually their behaviours, competitive or cooperative. Competitive interactions are typically associated with cross-immunity, such that one individual that has been infected with one disease has a lower infection rate for the other, although they can be related with non-pharmaceutical interventions such as isolation. In contrast, cooperative effects represent the weakening of individuals, such that they are more likely to acquire secondary infections, commonly known as co-infection; this is something that can also happen when the symptoms of both infections are hard to distinguish and the patients share space at the hospitals.

Several mathematical models analyse the effects of interactions between pairs of infections [1,2]. In these models, individuals that have been infected by one disease display a modified infection rate with respect to the other disease, such that the infection rates of infected or recovered individuals are different from those of susceptible individuals. In fact, interactions between infections have been reported for pairs such as leprosy-tuberculosis [3], tuberculosis-HIV [4], hepatitis-HIV [5], or zika and dengue [6].

We model the interaction between the spreading processes of COVID-19 and seasonal influenza. The independent COVID-19 infection is modelled as a SEIR (Susceptible-Exposed-Infected-Recovered) dynamics, such that susceptible individuals in contact with infected individuals become exposed with a rate $\beta_c$, spontaneously become infected with a rate $1/\tau_{Ec}$ and recover from the infected state with a rate $1/\tau_c$. The dynamics of independently spreading influenza is described by the SIR (Susceptible-Infected-Recovered) model, analogous to that of COVID-19 but without the exposed state, with an infection rate $\beta_i$ and a recovery rate $1/\tau_i$. We consider the deterministic mean-field equations describing these processes, extracting the parameters from the literature (Table I, [7,8]). The interaction implies that individuals that are not susceptible to COVID-19 (influenza) become infected with influenza (COVID-19) with a rate $k_i \beta_i$ ($k_c \beta_c$) (Fig. 1). Values of $k_i$ and $k_c$ higher than 1 represent a cooperative interaction (known as co-infection), while values that are lower than 1 describe competitive interactions, typically introduced as cross-immunity, although this can also illustrate the non-pharmaceutical intervention of keeping infected individuals in quarantine and, hence, isolating them from the other pathogen.

The dynamics describing the temporal evolution of the 12 states introduced in Fig. 1 is simplified in the dynamics of 8 variables: $s = [S]$, $e_c = [E_c] + [E_c\ I_i] + [E_c\ R_i]$, $q_c = [I_c] + [I_c\ I_i] + [I_c\ R_i]$, $t_c = [E_c] + [I_c] + [R_c]$, $f_c = [R_c] + [R_c\ I_i] + [R_c\ R_i]$, $q_i = [I_i] + [E_c\ I_i] + [I_c\ I_i] + [R_c\ I_i]$, $t_i = [I_i] + [R_i]$, and $f_i = [R_i] + [E_c\ R_i] + [I_c\ R_i] + [R_c\ R_i]$, where the brackets indicate the density of individuals in that state. These variables evolve according to the following system of ordinary differential equations:

$$\frac{ds}{dt} = -\beta_c s q_c - \beta_i s q_i$$
$$\frac{de_c}{dt} = \beta_c s q_c + k_c \beta_c t_i q_c - \tau_{Ec}^{-1} e_c$$
$$\frac{dq_c}{dt} = \tau_{Ec}^{-1} e_c - \tau_c^{-1} q_c$$
$$\frac{dt_c}{dt} = \beta_c s q_c - k_i \beta_i t_c q_i \quad (1)$$
$$\frac{df_c}{dt} = \tau_c^{-1} q_c$$
$$\frac{dq_i}{dt} = \beta_i s q_i + k_i \beta_i t_c q_i - \tau_i^{-1} q_i$$
$$\frac{dt_i}{dt} = \beta_i s q_i - k_c \beta_c t_i q_c$$
$$\frac{df_i}{dt} = \tau_i^{-1} q_i$$

We numerically integrate the equations describing this dynamics, starting from the initial condition $s(t=0) = 1-\epsilon$, $q_c(t=0) = t_c(t=0) = q_i(t=0) = t_i(t=0) = \epsilon/2$, $e_c(t=0) = f_c(t=0) = f_i(t=0) = 0$, with $\epsilon = 0.01$.

First, we analyse the non-interacting scenario ($k_i = k_c = 1$), observing that the dynamics of COVID-19 reaches a higher peak before the peak of influenza (Fig. 2). Specifically, it takes influenza twice the time that it takes COVID-19 to reach its peak, with the prevalence at the peak of COVID-19 being 5 times higher than that displayed at the peak of influenza. We observe that, under this mean-field dynamics, the COVID-19 reaches all the population along with this infection wave (cumulative prevalence $f_c$ higher than 98%), while influenza reaches a cumulative prevalence $f_i$ of 44% (see Supplementary Text 1, Fig. 2 inset), such that under the mean-field well-mixed assumption, the cumulative prevalence of people with both infections approaches $f_i$. The simplification of our model of 12 states into a set of 8 differential equations does not allow us to isolate the number of individuals affected by at least one active infection, but we approximate it by $Q = q_c + q_i$, which is a maximum bound of it and a good estimator as far as [$I_c$ $I_i$] is low. This parameter reaches its peak close to the peak of COVID-19, and displays a long tail due to the slower spreading of influenza (Fig. 2).

In the interactive scenarios, the COVID-19 infection dominates the dynamics, and its prevalence curve does not change considerably due to the interactions, but it influences that of influenza (Fig. 3). Specifically, when the interaction is competitive for the infection of COVID-19 ($k_c < 1$), the time at which the maximum prevalence of COVID-19 is reached does not significantly change (variations smaller than 0.2 days), and it has a slightly lower prevalence (relative decrease in 2% of $k_i$=0.5 and 4% for $k_i$=2, Fig. 3a,b), with the opposite behaviour for cooperative infections (relative increase in 2% for $k_i$=0.5 and 3% for $k_i$=2, Fig. 3c,d), but without experiencing major changes in the peak prevalence. In contrast, competitive interactions in the influenza infection ($k_i < 1$) considerably reduce the peak (relative decrease in 68% for both $k_c$=0.5 and $k_c$=2) and the time taken to reach it (from 41.60 days to 15.40 and 15.15 days for $k_c$=0.5 and $k_c$=2, respectively, Fig. 3a,c), because as the COVID-19 infection initially grows faster than the influenza infection, the competitive interaction flattens the prevalence curve of influenza. The cooperative interaction for influenza infection ($k_i > 1$) also reduces the time taken to reach its maximum prevalence (28.91 and 28.68 days for $k_c$=0.5 and $k_c$=2, respectively) but increases the height of this peak (relative increases in 500% and 509% for $k_c$=0.5 and $k_c$=2, respectively, Fig. 3b,d). When we consider the variable $Q = q_c + q_i$, which is a good estimator of the demand in the health care centres, we realize that first, the temporal extent of the dynamics is halved, from approximately 100 to 50 days, and that this demand is affected by $k_i$, such that $k_i = 0.5$ slightly reduces the peak demand and $k_i = 2$ increases it in a factor 1.5 (Fig. 3 insets).

Interaction between the infections of COVID-19 and influenza is translated into a synchronization of their temporal behaviours, driven by the COVID-19 infection, which initially grows faster (Fig. 2). We explore the prevalence at the peaks $q_M$ and their arrival times $T_M$ in two cases: an asymmetric interaction ($k_c = 1$, $k_i$ variable), and a symmetric interaction ($k_c = k_i$ variable). In the asymmetric scenario, any interaction ($k_i > 1$), decreases the arrival time of the influenza peak (Fig. 4a). Interestingly, this interaction makes the prevalence of influenza higher than that of COVID-19, and it reaches a maximum for $k_i = 5.8$, with also the influenza peak arriving earlier than that of COVID-19, such that the influenza prevalence curve could be interpreted as an early-warning of the COVID-19 curve, arriving later due to the exposure period. In the symmetric interaction, we find a similar behaviour for influenza, while COVID-19 does not substantially modify the arrival time of its peak, and there is more variation on its peak intensity, even if the change of values is much lower than the observed for influenza (Fig. 4b). We observe an optimal value for the interaction parameter $k_i$ in both cases such that the peak of influenza displays a maximum prevalence, interpreted as a resonance between both infection processes, such that if $k_i$ is higher, the peak of influenza will occur earlier, but in a moment where the fraction of exposed individuals to COVID-19 is not maximum yet.

Mathematical modelling of infectious diseases is becoming a fundamental tool for forecasting and policy-making, leading to a remarkable effort for revealing the statistical properties of the models of spreading dynamics [9]. In fact, the interaction between multiple infections represents an additional degree of complexity. Our model illustrates the interaction between two infections described by different dynamics in their independent spread. We recognize that this work has some limitations. First, it considers a mean-field dynamics without focusing on the interaction network, with details leading to a rich dynamics in the frame of interactive infections in static complex networks [10] or temporal networks [11]. Additionally, we consider that the interaction modifies the infection rates, while an extended model would include the effects of the interaction in recovery rates, as observations reflect that coinfection of SARS-CoV-2 and other pathogens leads to longer lengths of the hospitalary stays [12].

Research on the interactions of COVID-19 with other infections is still ongoing. Indeed, the competitive process due to the cross-immunity is controversial, as the immune system displays different response pathways to COVID-19 and influenza [13], in contrast to recent reports that revealed that T cells related to the immune response to common cold coronaviruses react against SARS-CoV-2 [14,15]. Regarding coinfection of SARS-CoV-2 and other respiratory pathogens, empirical observations reported a reduced frequency of other pathogens in individuals that were tested positive for SARS-CoV-2 (20.7%) in contrast to the frequency in those tested negative (26.7%) [16], while other observations reported even lower frequencies of other respiratory pathogens, represented by 2.99% for individuals tested positive and 13.1% for individuals tested negative [17].

Apart from the biomedical interpretation of the interactions, our well-mixed framework may include the effects of non-pharmaceutical interventions, leading to a reduced prevalence of other diseases, such as influenza, due to the preventive lockdown against the spreading of COVID-19 [18,19], the slowing down of vaccination campaigns against other infections, thus facilitating their spread [20], or the disruption on treatment and preventive actions against HIV [21].

We have shown a synchronization process between COVID-19 and influenza. In fact, correlations between the empirical number of positive-tested patients of COVID-19 and excess influenza-like illness cases have been reported [22,23], which could emerge either from considering non-positive COVID-19 patients within those cases or from the synchronization between COVID-19 and influenza-like illness outbreaks. Our results reveal how the prevalence information from other infections may inform us about the evolution of COVID-19 outbreaks and, at the same time, facilitate the inference of the prevalence curve when the reporting managers are not able to distinguish between the symptoms of the considered interacting infections.

**Tables**

| Infection | Infection rate (days$^{-1}$) | Exposure time (days) | Infection time (days) | Ref. |
|---|---|---|---|---|
| COVID-19 | $\beta_c$ = 1.12 | $\tau_{Ec}$ = 3.69 | $\tau_c$ = 3.47 | [7] |
| Seasonal influenza | $\beta_i$ = 0.32 | x | $\tau_i$ = 4.1 | [8] |

Table I. Parameters extracted from the literature for the dynamics of COVID-19 and seasonal influenza.

**Figures**

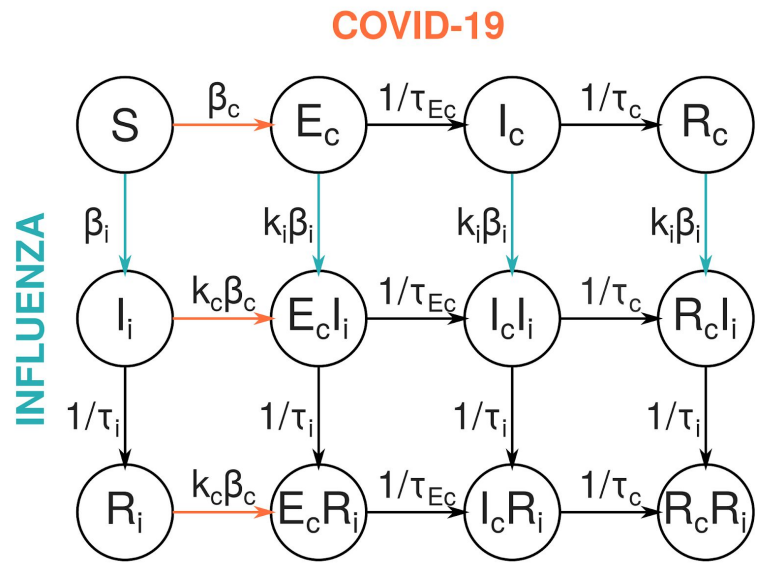

**Fig. 1.** Scheme of the dynamics. The letters inside the circles denote the 12 possible states in the SEIR-SIR dynamics, the letters over the arrows represent the transition rates between the states, and the arrow colors depict the exposure to other states necessary for each transition, with black, orange and blue colors representing, respectively, spontaneous processes, processes under the exposure to the COVID-19 infected population, and processes under the exposure to the population infected with influenza.

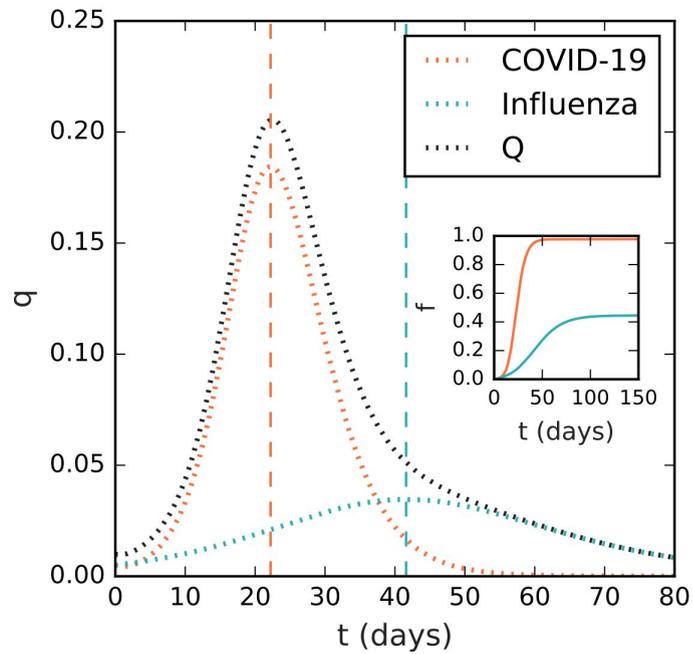

**Fig. 2.** Temporal series of the fraction of infected individuals q for COVID-19 and influenza in the case of non-interacting infections, i.e. $k_c = k_i = 1$. The dashed lines represent the peak of infected individuals for COVID-19 (orange), influenza (blue) and both ($Q = q_c + q_i$, black). Inset: temporal series of the cumulative prevalence f of COVID-19 and influenza.

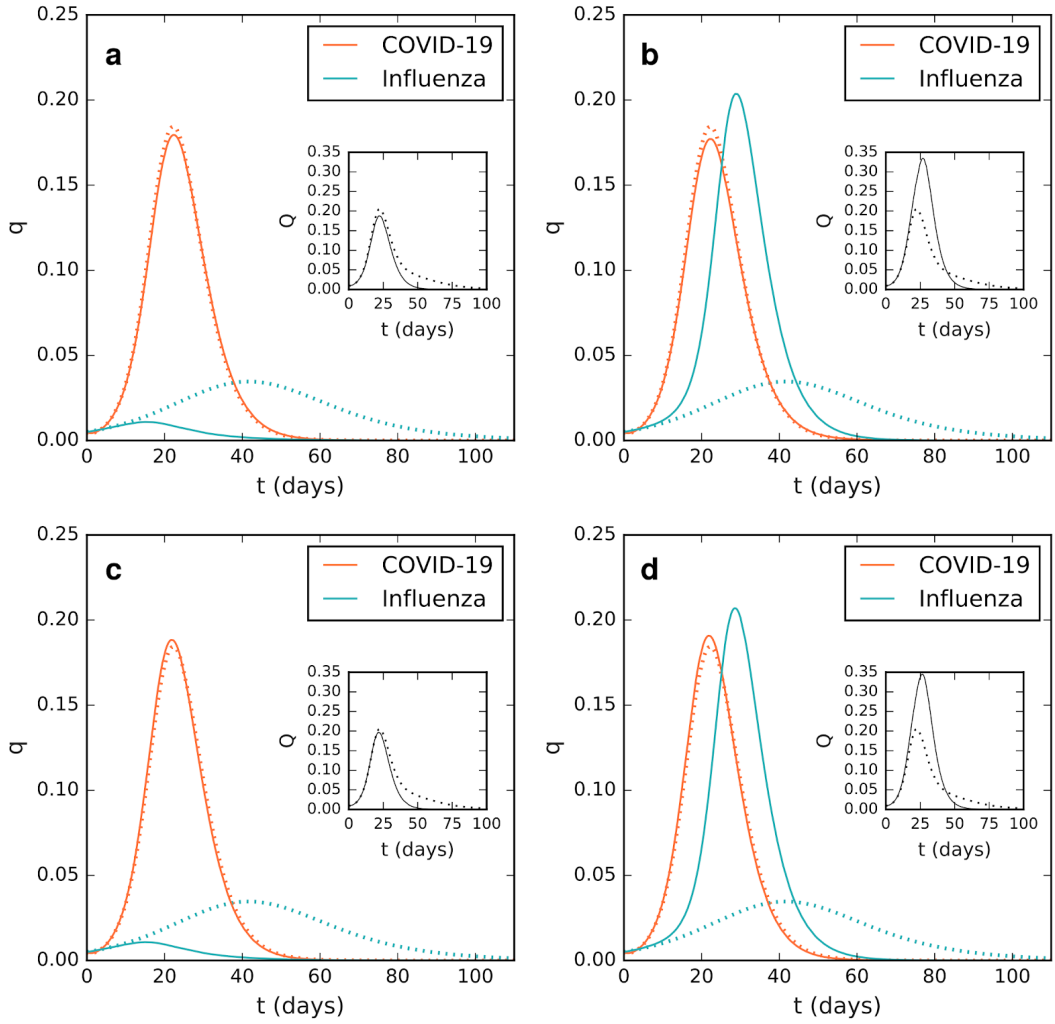

**Fig. 3.** Temporal series of the fraction of infected individuals q for COVID-19 and influenza in the interactive scenarios **a,** both experience competitive effects $k_c = k_i = 0.5$, **b,** $k_c = 0.5$, $k_i = 2$, **c,** $k_c = 2$, $k_i = 0.5$, **d,** both experience cooperative effects $k_c = k_i = 2$. Insets: fraction of individuals $Q = q_c + q_i$ affected by at least one infection. Dotted lines represent the case $k_c = k_i = 1$, while continuous lines correspond to each case in **a-d**.

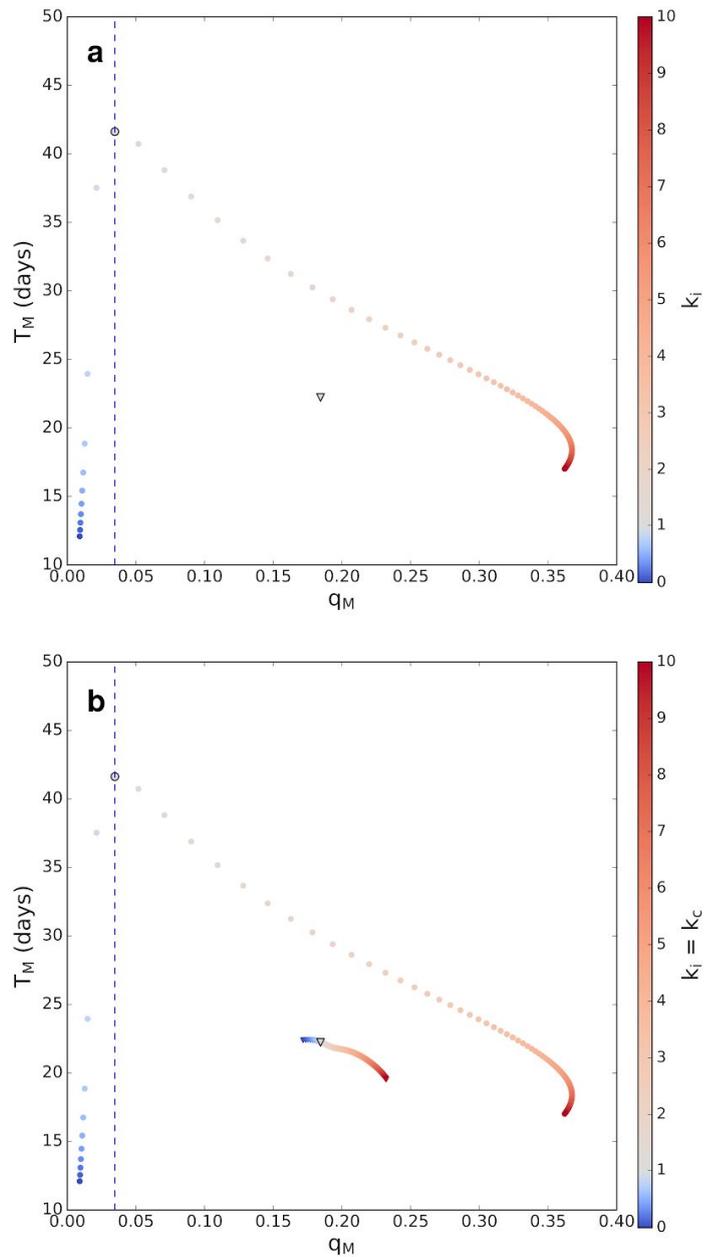

**Fig 4.** Phase diagram of the prevalence at the peak $q_M$ and the time $T_M$ at which the peak is reached, for different interacting scenarios. **a,** $k_c = 1$, varying $k_i$ and **b,** varying the interaction strengths $k_c$ and $k_i$ under the constraint $k_c = k_i$. The vertical dashed line is the analytical prediction for the maximum prevalence of influenza with $k_i = 1$ (see Supplementary Text 2). Circles and triangles represent, respectively, the values associated with influenza and COVID-19, colors indicate the interaction strength and the symbols with black perimeters correspond to the case $k_c = k_i = 1$.

# Coupling between COVID-19 and seasonal influenza leads to synchronization of their dynamics
## Supplementary material

**Supplementary Text 1**

Considering a Susceptible-Infected-Recovered (SIR) dynamics, described by

$$\frac{ds}{dt} = -\beta s i$$
$$\frac{di}{dt} = \beta s i - \tau^{-1} i$$
$$\frac{dr}{dt} = \tau^{-1} i$$

we obtain $\frac{ds}{dr} = -\beta \tau s$ which, taking into account $s(t=0) = s_0$ and $r(t=0) = 0$, leads to $s(t) = s_0 \, e^{-\beta \tau \, r(t)}$, and considering $s(t \to \infty) + r(t \to \infty) = 1$, we obtain the non-linear equation $r(t \to \infty) = 1 - s_0 \, e^{-\beta \tau \, r(t \to \infty)}$.

Solving numerically this equation with the parameters $s_0 = 1 - \epsilon/2$, $\epsilon = 0.01$, $\beta = 0.32$ and $\tau = 4.1$, we obtain $r(t \to \infty) \approx 0.445$.

**Supplementary Text 2**

Considering a SIR model, with the evolution described in Supplementary Text 1, the maximum value of the fraction of infected is obtained imposing $di/dt = 0$, which leads to $s_p = 1/\beta\tau$, where the subindex p indicates the value at the peak. Considering $s(t) = s_0 \, e^{-\beta \tau \, r(t)}$ and $s_p + i_p + r_p = 1$, we obtain $i_p = 1 - 1/R(1 + \log s_0/R)$, where $R = 1/\beta\tau$ is the basic reproductive number.

Substituting $s_0 = 1 - \epsilon/2$, with $\epsilon = 0.01$ and the parameters detailed in Table I, we obtain $i_p \approx 0.0346$.